\documentstyle[prl,preprint,aps,psfig,amstex]{revtex}
\newcommand{\rar}{\rightarrow}
\newcommand{\al}{\alpha}
\newcommand{\be}{\beta}
\begin{document}
\preprint{M\'exico ICN-UNAM 98-05}
\title{ $H_3^{++}$  molecular ions can exist in strong magnetic
fields}
\author{Alexander Turbiner\cite{byline}\cite{byline2}, Juan Carlos Lopez V. 
and Ulises Solis H.}
\address{Instituto de Ciencias Nucleares, UNAM, Apartado Postal 70-543,
        04510 M\'exico D.F., M\'exico}

\maketitle

\smallskip

\centerline{\it Dedicated to the memory of B.B.~Kadomtsev}

\begin{abstract}
Using a variational method it is shown that for magnetic fields
$B > 10^{11}$ G there can exist a molecular ion $H_3^{++}$.
\end{abstract}



Many years ago the formation of unusual chemical compounds in the
 presence of a strong magnetic field whose existence is impossible
 without strong magnetic fields was predicted by Kadomtsev and
 Kudryavtsev \cite{Kadomtsev:1971} and Ruderman \cite{Ruderman:1971}
 (for a recent advances and a review, see, for example,
 \cite{Salpeter:1995,Salpeter:1997} and references therein).  In
 particular, using a semi-classical analysis it was shown that the
 influence of the strong magnetic field leads to the appearance of
 linear molecules (linear chains) situated along magnetic lines. The
 transverse size of such systems should be of the order of the
 cyclotron radius ${\hat\rho} \sim B^{-1/2}$ (a.u.), while the
 longitudinal size remains of a molecular (atomic) order. These
 systems are called needle-like. An important consequence of such a
 quasi-one-dimensionality of Coulombic systems is the possibility of
 effectively compensating the Coulombic repulsion of nuclei.

The goal of this Letter is to present the first quantitative study of
the molecular ion $H_3^{++}$ in a strong magnetic field which provides
theoretical evidence that such a system can exist in magnetic field
$B >  10^{11}$G.  Our study is limited to an exploration of the
ground state.  Throughout the present work it is assumed that the
Born-Oppenheimer approximation holds which implies that the positions
of protons are fixed. Exactly as for $H_2^+$ the configuration we
consider corresponds to the case when the three protons are aligned
with the magnetic field (linear chain, see above). Spin effects
(linear Zeeman effect) are neglected. The magnetic field ranges from 0
up to $10^{13}$ G, where it is assumed that a contribution of 
relativistic corrections can still be neglected (for a discussion see,
for instance, \cite{Salpeter:1997} and references therein). Finally,
it is also demonstrated that the molecular ion $H_2^{+}$ is the most
bound one-electron molecular system in a constant magnetic field.

The present calculation is carried out in the framework of a
variational method using a {\it unique} simple trial function equally
applicable to any value of the magnetic field strength.  Very recently
\cite{Lopez:1997}, this strategy was successfully applied to study the
ground state of the molecular ion $H_2^+$ and a simple 10-parameter
trial function allowed one to get the best (lowest) values of the
ground state energy for magnetic fields from 0 up to $10^{13}$ G
(except $B=0$, where the relative accuracy was about $10^{-5}$ in
comparison with the best calculations). It turned out that it was not
only the best calculation in the region of accessible magnetic fields
$0-10^{13}$G but the only calculation which tackled the problem using
a single approach.  Since the key point of a successful study is a
wise choice of trial functions, we give a detailed description of how
to choose trial functions adequate to the problem in hand.

A constructive criterion for an adequate choice of trial function was
formulated in \cite{Turbiner:1980} and further development was
presented in \cite{Turbiner:1984,Turbiner:1987}. In its simplest form
the criterion is the following. The trial function $\Psi_t(x)$ should
contain all symmetry properties of the problem in hand. If the ground
state is studied, the trial function should not vanish inside the
domain where the problem is defined. The potential
$V_t(x)=\frac{{\mathbf\nabla}^2 \Psi_t}{\Psi_t}$, for which the trial
function is an eigenfunction, should reproduce the original potential
near singularities as well as its asymptotic behavior. The use of this
simplest possible recipe has led to a unique one-parameter trial
function, which in particular, made it possible to carry out the first
qualitative study of the ground state of the hydrogen molecule $H_2$
in the region of both weak and strong magnetic fields
\cite{Turbiner:1983}.  Later a few-parameter trial function was
proposed for a description of the hydrogen atom in an arbitrary
magnetic field, which led, for the low-excited states, to an accuracy
comparable with the best calculations
\cite{Turbiner:1987,Turbiner:1989}.

Now we wish to apply the above  recipe to the ion $H_3^{++}$.
Let us first introduce notation (see Fig.1).  We consider three
attractive identical centers of unit charge  situated on the $z$-axis at
origin and at a distance $R_-,R_+$ from the origin, respectively.  The
magnetic field of strength $B$ is directed along the $z$ axis and
$r_{1,2,3}$ are the distances from the electron to the first (second,
third) center, respectively. The quantity $\rho$ is the distance from
the electron to the $z$-axis. Through the paper the Rydberg is used as
the energy unit. For the other quantities standard atomic units are
used. The potential corresponding to the problem we study is given by
\begin{equation}
\label{e1}
V=\frac{2}{R_-} + \frac{2}{R_+} + \frac{2}{R_-+R_+}
-\frac{2}{r_1} - \frac{2}{r_2} - \frac{2}{r_3} +
\frac{B^2 \rho^2}{4},
\end{equation}
where the first three terms have the meaning of the classical Coulomb
energy of interaction of three charged centers. The recipe dictates
that the trial functions should behave in a Coulomb-like way near the
centers, correspond to two-dimensional oscillator behavior in the
$(x,y)$ plane at large distances and be permutationally-symmetric with
respect to exchange of positions of the centers. It seems quite
natural that the equilibrium configuration corresponding to minimal
total energy of the system should appear at $R_-=R_+$.

 One of the simplest functions satisfying the above recipe is the
Heitler-London type function multiplied by the lowest Landau orbital:
\begin{equation}
\label{e2}
\Psi_1= {e}^{-\al_1  (r_1 + r_2 + r_3) - \be_1 B \rho^2 /4},
\end{equation}
(cf. Eq. (2.2) in \cite{Lopez:1997}), where $\al_1,\be_1$ are
variational parameters. It has a total of four  variational parameters if
the internuclear distances $R_-,R_+$ are taken as parameters. It is
quite natural from a physical viewpoint to assume that a function of
the Heitler-London type gives an adequate description of the system
near the equilibrium position. The potential
$V_1(x)=\frac{{\mathbf\nabla}^2\Psi_1}{\Psi_1}$, corresponding to this
function is:
\begin{eqnarray}\label{e3}
V_1 \ &=&\ 3 \al^2_1 -B \be_1 -2 \al_1
\Bigg(\frac{1}{r_1}+\frac{1}{r_2}+\frac{1}{r_3}\Bigg)+\frac{\be^2_1
B^2\rho^2}{4}+ 2\al^2_1
\Bigg[
   \frac{1}{r_1r_2}\Big(\rho^2+z(z+R_+)\Big)+
   \nonumber\\ &&
   \frac{1}{r_2r_3}\Big(\rho^2 + z(z-R_-)\Big)+
   \frac{1}{r_1r_3}\Big(\rho^2 + (z-R_-)(z+R_+)\Big)
\Bigg]
+\ \al_1  \be_1 B\rho^2
\Bigg(\frac{1}{r_1}+\frac{1}{r_2}+\frac{1}{r_3}\Bigg)
\end{eqnarray}
It is clear that this potential reproduces the original potential (\ref{e1})
near Coulomb singularities  as well as at large distances, $|x,y|
\rar \infty$.

The Hund-Mulliken-type function multiplied by the lowest Landau orbital
is another possible trial function:
\begin{equation}
\label{e4}
\Psi_2= \Big( {e}^{-\al_2  r_1} + e^{-\al_2  r_2} +
e^{-\al_2  r_3}\Big) {e}^{-\be_2 B \rho^2 /4} \ ,
\end{equation}
(cf. Eq. (2.4) in \cite{Lopez:1997}), where $\al_2,\be_2$ are
variational parameters. It is obvious that this function, in the
absence of a magnetic field, gives an essential contribution to a
description of the region of large internuclear distances. The
calculations we have performed show that this property remains valid
for all magnetic fields up to $10^{13}$ G. Like Eq. (\ref{e2}), the
trial function (4) is characterized by four variational parameters. This
function, when both internuclear distances are large, corresponds to a
decay $H_3^{++} \rar H + p + p$.

Another trial function supposedly describes a decay mode $H_3^{++} \rar
H_2^{+} + p$ and could be taken to be of the form
\begin{equation}
\label{e4-2}
\Psi_3= \Big( {e}^{-\al_3  (r_1+r_2)} + e^{-\al_3  (r_1+r_3)} +
e^{-\al_3  (r_2+r_3)}\Big) {e}^{-\be_3 B \rho^2 /4} \ ,
\end{equation}
where $\al_3,\be_3$ are variational parameters. Finally, it will
become obvious that the function (\ref{e4-2}) does give the dominant
contribution to the large internuclear distances. Eq. (\ref{e4-2})
also depends on four variational parameters.

To take into account both equilibrium and large distances, we
use an interpolation of Eqs. (2), (4) and (5). There are three natural
approaches to interpolate:
\begin{itemize}
\item[(i)] a total non-linear superposition:
\begin{eqnarray}\label{e5}
\Psi_{4-{nls-t}} &=& \Big(
 {e}^{ -\al_4  r_1 - \al_5  r_2 - \al_6  r_3} +
 {e}^{ -\al_5  r_1 - \al_4  r_2 - \al_6  r_3} +
 {e}^{ -\al_4  r_1 - \al_6  r_2 - \al_5  r_3} +
\nonumber \\ &&
 {e}^{ -\al_6  r_1 - \al_4  r_2 - \al_5  r_3} +
 {e}^{ -\al_5  r_1 - \al_6  r_2 - \al_4  r_3} +
 {e}^{ -\al_6  r_1 - \al_5  r_2 - \al_4  r_3}
 \Big) {e}^{ -\be_4 B\rho^2/4}\ ,
\end{eqnarray}
(cf. Eq. (2.5) in \cite{Lopez:1997}), where $\al_{4,5,6}, \be_4$ are
variational parameters. The function (\ref{e5}) is a three-center
modification of the Guillemin-Zener type function used for the
description of the molecular ion $H^+_2$ in a magnetic field
\cite{Lopez:1997}.  If $\al_{4}=\al_{5}=\al_{6}\equiv \al_{1} $, the
function (\ref{e5}) reduces to Eq. (2).  When $\al_{4}\equiv \al_{2},
\al_{5}=\al_{6}=0$, it coincides with Eq. (4).  Finally, if
$\al_{4}=\al_{5}\equiv \al_{3}, \al_{6}=0$, the function (\ref{e5})
reduces to Eq. (\ref{e4-2}).  In total there are 6 variational
parameters characterizing the trial function (\ref{e5});

\item[(ii)] a partial non-linear superposition:\\
 this appears if in Eq. (\ref{e5}) the two parameters are equal,
 for instance, $\al_{4}=\al_{5}$:
\begin{equation}
\label{e6}
\Psi_{4-{nls-p}}= \Big(
 {e}^{-\al_4  (r_1 + r_2) - \al_6  r_3} +
 {e}^{-\al_4  (r_1 + r_3) - \al_6  r_2 } +
 {e}^{-\al_4  (r_2 + r_3) - \al_6  r_1 }
 \Big) {e}^{-\be_4 B\rho^2/4}\ ,
\end{equation}
This function can be considered as a non-linear interpolation between
Eqs. (4) and (5).

\item[(iii)] a linear superposition of Eqs. (2), (4), (5)
\begin{equation}
\label{e7}
\Psi_{4-{ls}}= A_1 \Psi_1 + A_2 \Psi_2 + A_3 \Psi_3 \ ,
\end{equation}
where the relative weights of Eqs. (2), (4), (5) in Eq. (7) are taken as extra
variational parameters. This is a 10-parameter trial function.

\end{itemize}
 Of course, as a natural continuation of the above
interpolation procedure one can take a linear superposition of all five
functions (2), (4), (5), (6), (7):
\begin{eqnarray}\label{e8}
\Psi_5 &=& A_{4-{nls-t}}\Psi_{4-{nls-t}}+A_{4-{nls-p}}\Psi_{4-{nls-p}}+
 A_{4-{ls}}\Psi_{4-{ls}} 
\nonumber\\
&=&
 A_{4-{nls-t}}\Psi_{4-{nls-t}}+ A_{4-{nls-p}}\Psi_{4-{nls-p}}+ A_1 \Psi_1
 + A_2 \Psi_2 +A_3 \Psi_3  \ ,
\end{eqnarray}
(cf. Eq. (2.7) in \cite{Lopez:1997}), where again, as in
the case of the function (8) the relative weights of different,
`primary' trial functions are considered as variational parameters. In
total, the trial function (9) is characterized by 17 variational
parameters. However, only part of our calculation is carried out using
this function. Usually, some particular cases of Eq. (9) are
explored. The general case will be presented elsewhere. The
minimization procedure is carried out using the standard minimization
package MINUIT from CERN-LIB on a Pentium-Pro PC. All integrals were
calculated using the CERN-LIB routine DGAUSS with relative accuracy $\le
10^{-7}$.

In Table I the results of our variational calculations are
presented. It is quite remarkable that for magnetic field strengths
$ > 10^{11}$ G there exists a minimum of total energy in the
$(R_{+},R_{-})$ plane.  Furthermore, for such magnetic fields the
value of the energy at the minimum correspondent to the total energy
of $H_3^{++}$ is always lower than the total energy of the hydrogen
atom but higher than that of $H_2^{+}$. Hence the decay mode $H_3^{++}
\rar H+p+p$ is forbidden. However, $H_3^{++}$ is unstable with respect
to the decay $H_3^{++} \rar H_2^+ + p$. It seems natural to assume
that even if one-electron systems like $H_4^{+++}, H_5^{++++}$
etc. would be bounded, their total energies will be larger than the
total energy for $H_3^{++}$. This assumption and comparison of the
total energies of $H, H_2^+, H_3^{++}$ (see Table I) allows one to
conclude that $H_2^{+}$ is the most stable one-electron system in a
constant magnetic field oriented along the magnetic field. The
equilibrium distances for $H_3^{++}$ decrease with the growth of the
magnetic field: the ion $H_3^{++}$, like $H_2^{+}$, becomes more and
more compact. It is worth noting that for both $H_2^+$ and $H_3^{++}$
the average value $\langle z \rangle$ is much smaller than a `natural'
size of a system determined by the positions of the centers: $R_{eq}$
for $H_2^+$ and $({R_+}_{eq} + {R_-}_{eq})$ for $H_3^{++}$ (see
Fig. 1). In other words this means that the localization length of
electron is much smaller than the `natural' size of the system.

Fig.2 shows the electronic density distribution as a function of
magnetic field.  For a magnetic field $B\simeq 10^{11}$ G the
distribution has three clear maxima corresponding to the positions of
the centers, but the electron is situated preferably near the central
proton. The situation changes drastically with an increase of magnetic
field: the electron is localized near $z=0$, having almost no memory
of the two centers on either side. It is important to investigate
paths of possible tunneling. There are two explicitly pronounced
(symmetric with respect to $R_+ \leftrightarrow R_-$) valleys in the
electronic potential energy surfaces, $E_{total}(R_+,R_-)$ (see,
Fig.3a).

They vary from the position of the $H_3^{++}$ minimum to infinity
which corresponds to the $H_2^+ + p$ dissociation:
$({R_+}_{eq},{R_-}_{eq})\rar (\infty,R_{H_2^+}^{eq})$ and
$({R_+}_{eq},{R_-}_{eq}) \rar ( R_{H_2^+}^{eq},\infty)$, where
$R_{H_2^+}^{eq}$ is the equilibrium distance for the $H_2^+$ ion.  In
Fig.3b one can see the profile of the valley as a function of magnetic
field. Calculating the Gauss curvatures in the $H_3^{++}$ minimum one
can estimate the position of the ground state energy level and answer
the question of whether the well is deep enough to hold an energy
level.  It is always delicate to answer this question starting from
what `depth' of the well the level exists. Usually, it requires
special analysis.  We made an estimate and obtained the result that
for a magnetic field of $10^{11}$ G, the situation is not certain, the
well is probably still too shallow to hold the ground state energy
level. However, the well undoubtedly becomes sufficiently deep for
$10^{12-13}$ G. From the form of the profile (see Fig. 3b) it is quite
obvious that for $10^{12-13}$ G the barrier is rather high and the
probability of tunneling should be small.

\bigskip

The authors wish to thank K.G.~Boreskov (Moscow) for useful discussions.
A.T. thanks L.~Cederbaum (Heidelberg) for a valuable comment. Fruitful
discussions with P.O.~Hess (M\'exico) in the early stage of the work are
gratefully acknowledged.

This work is supported in part by DGAPA grant \linebreak \# IN105296
(M\'exico).

\begin{table}[hbt]
\begin{center} 
\begin{tabular}{lccccccccccc}
&\multicolumn{2}{c}{\( B=0 \) }&\multicolumn{3}{c}{\( B=10^{11} \) G}&\multicolumn{3}{c}{\( B=10^{12} \) G}&\multicolumn{3}{c}{\( B=10^{13} \) G}\\
&\( E \)&\( R_{eq} \)&\( E \)&\( R_{eq} \) &\(\langle z \rangle\) &\( E \)&\( R_{eq}
\) &\(\langle z \rangle\)&\( E \)&\( R_{eq}\)&\(\langle z \rangle\)\\
&\( (Ry) \)&\( (a.u.) \)&\( (Ry) \)&\( (a.u.) \) &\((a.u.)\) &\( (Ry) \)&\( (a.u.)
\) &\((a.u.)\)&\( (Ry) \)&\((a.u.)\)&\((a.u.)\)\\
\hline \\
\( H \)
   &-1.000 &-- & 36.929 &--  & &413.57&-- & &4231.6&-- & \\[3pt]
\( H_{2}^{+} \) 
   &-1.205 & 1.997  &35.036 &0.593&0.312 &408.300&0.283&0.174 
&4218.662&0.147&0.107\\[3pt]
\( H_{3}^{++} \)
   & --    & --       &36.429&0.803&0.432
&410.296&0.346&0.219&4220.090&0.165&0.121\\[3pt] 
\end{tabular}\par
\end{center}
\caption{ Data for the ground state of $H_3^{++}$ and a comparison
with data for other one-electron systems, $H$, $H_2^+$. Total energy
$E$ is in Rydbergs, the equilibrium distance $ R_{eq} \equiv {R_+}_{eq} =
{R_-}_{eq} $ (see text) and the average value of the longitudinal size
of the system $\langle z \rangle$ in $ a.u. $ Total energy for
hydrogen atom from [11]; data for $H_2^+$ from [5].}
\end{table}

\begin{figure}[tb]
\centerline{\hbox{
\psfig{figure=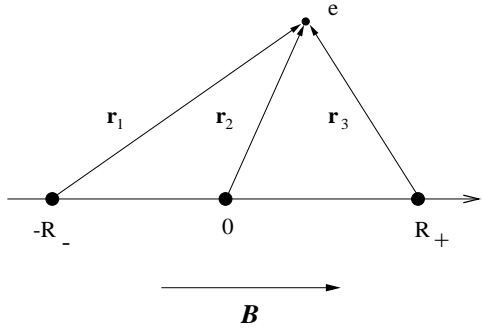,width=2.5in,angle=-90}
}}
\caption{$H_3^{++}$ in a magnetic field $B$. Explanation of the notation used}
\end{figure}
\begin{figure}[htb]
\[
\begin{array}{ccc}
{\psfig{figure=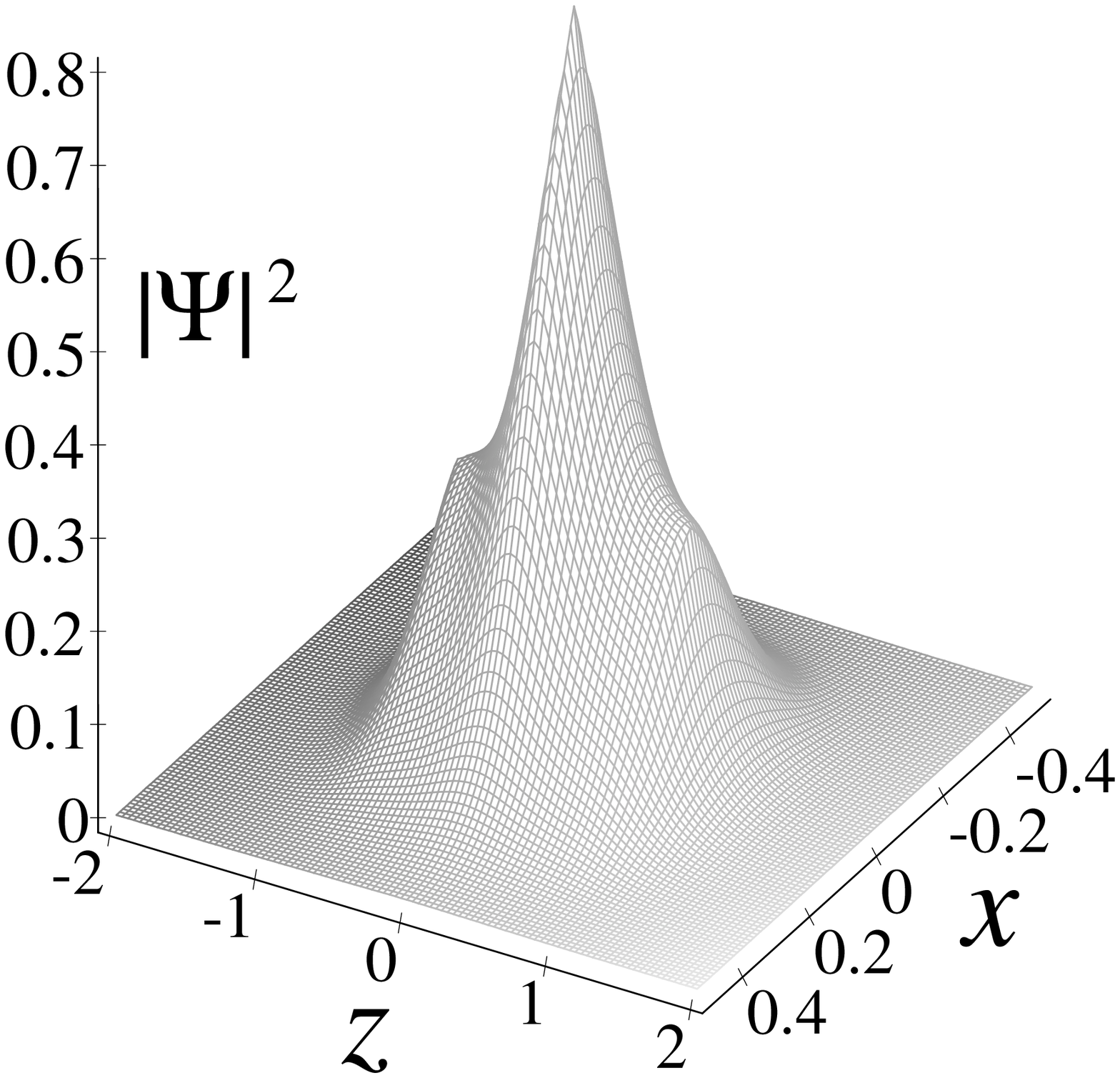,width=2.3in}} &
{\psfig{figure=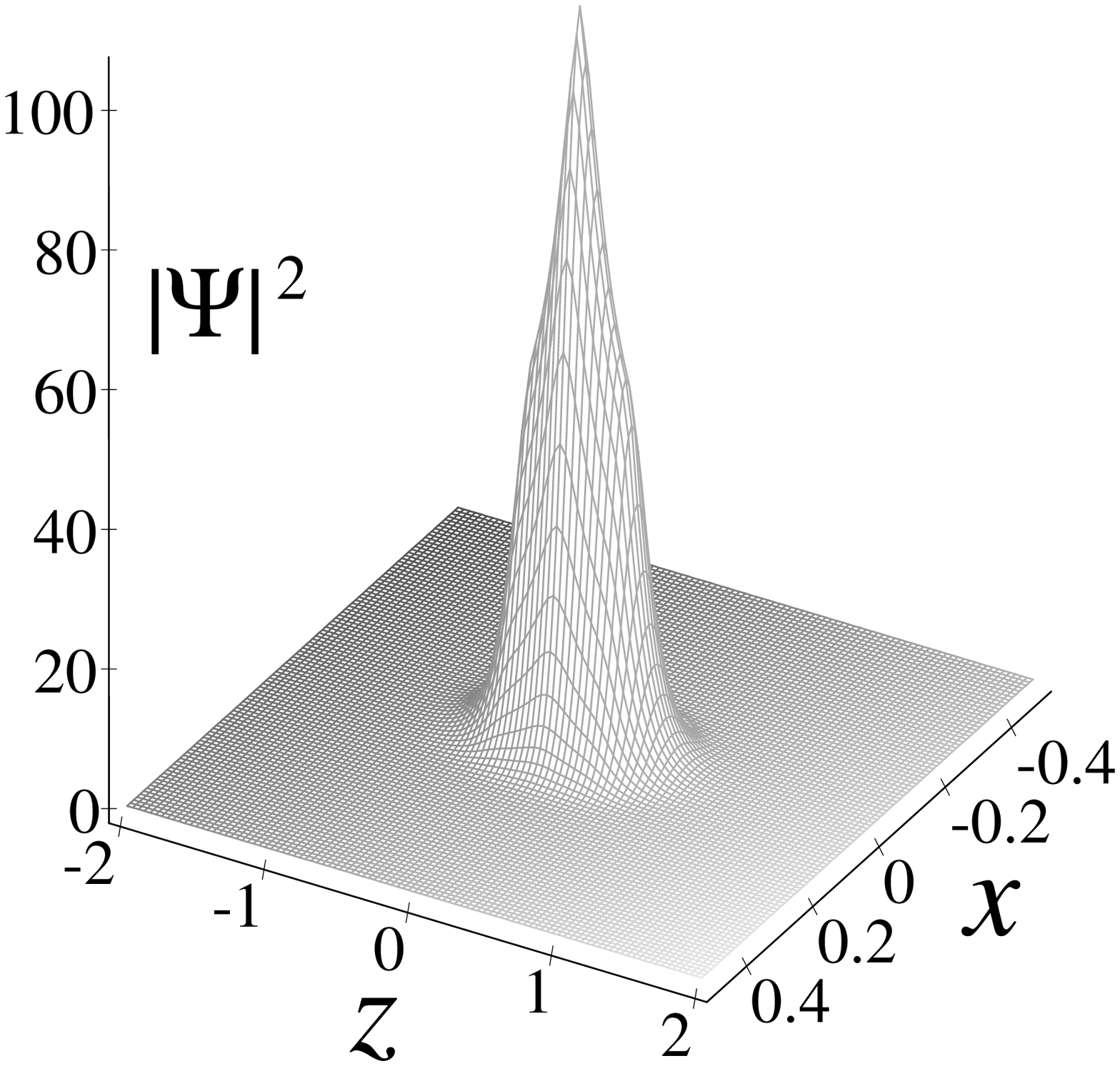,width=2.3in}} &
{\psfig{figure=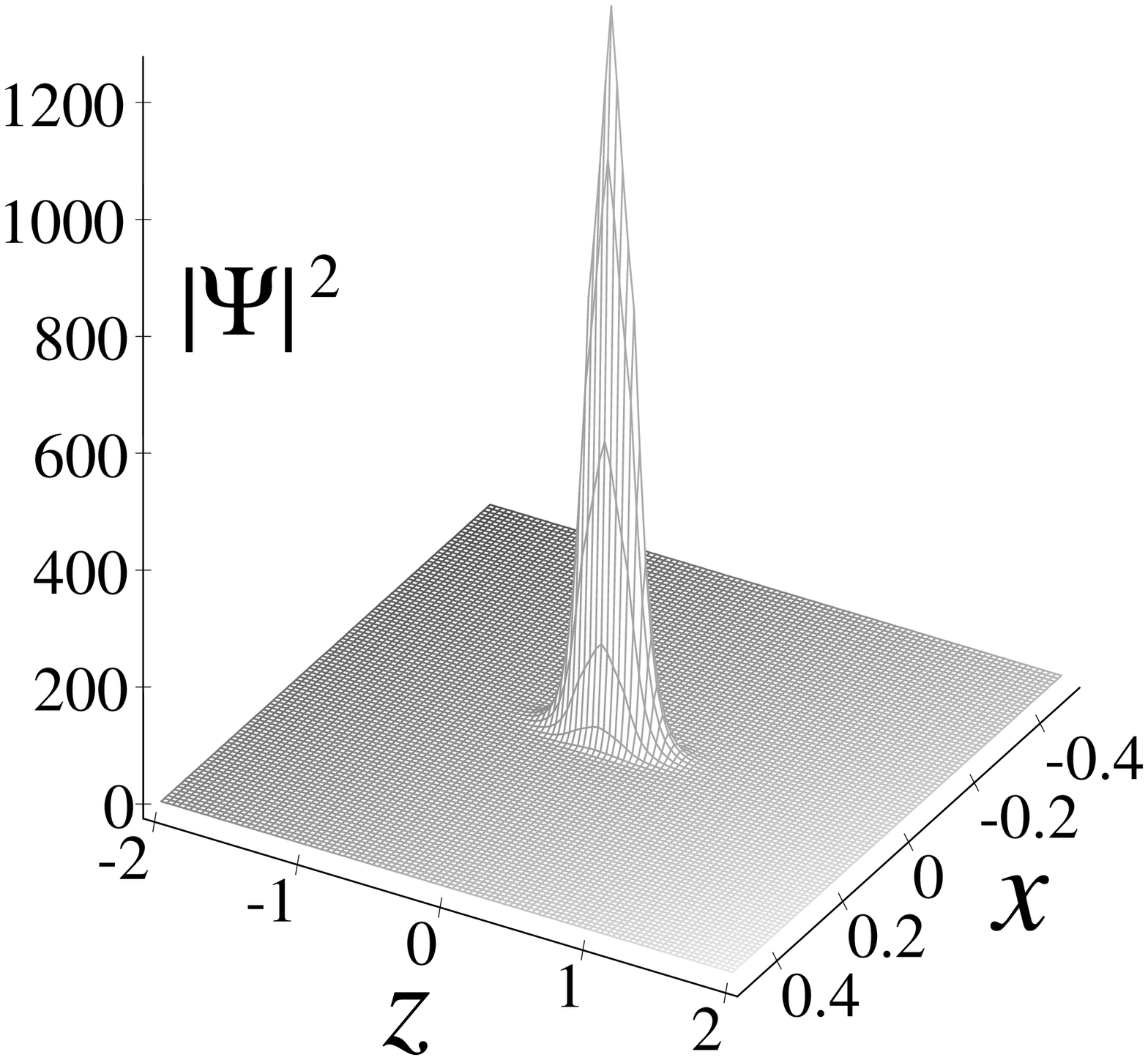,width=2.3in}} \\
(a) & (b) & (c)
\end{array}
\]
\caption{Electronic distribution for various magnetic fields:
$10^{11}$ G (a), $10^{12}$ G (b) and  $10^{13}$ G (c). It peaks more
and more sharply at origin with growth of magnetic field.}
\end{figure}

\begin{figure}[tb]
\[
\begin{array}{ccc}
{\psfig{figure=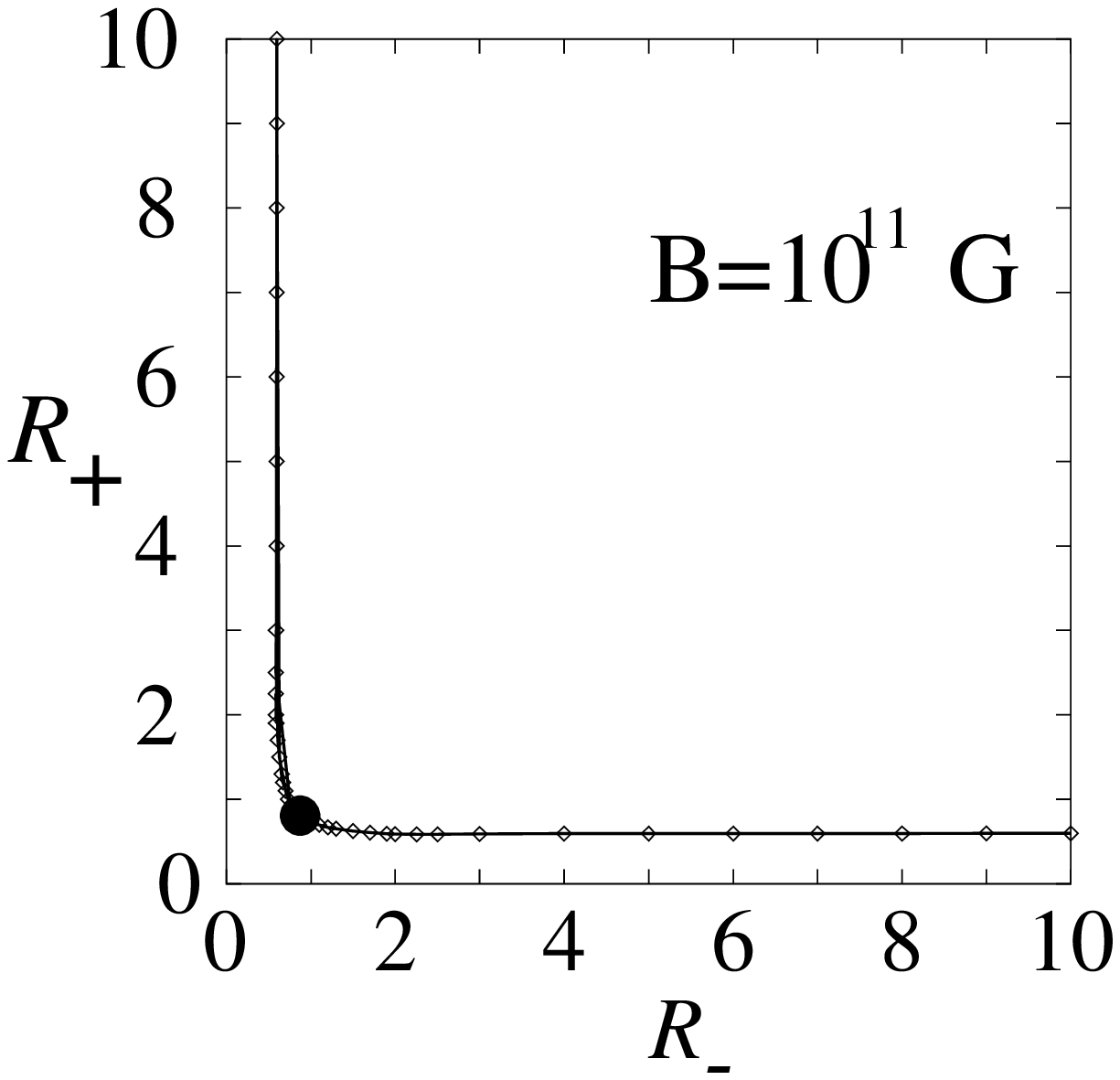,height=2.25in,angle=-90}} & \hspace{2cm}&
{\psfig{figure=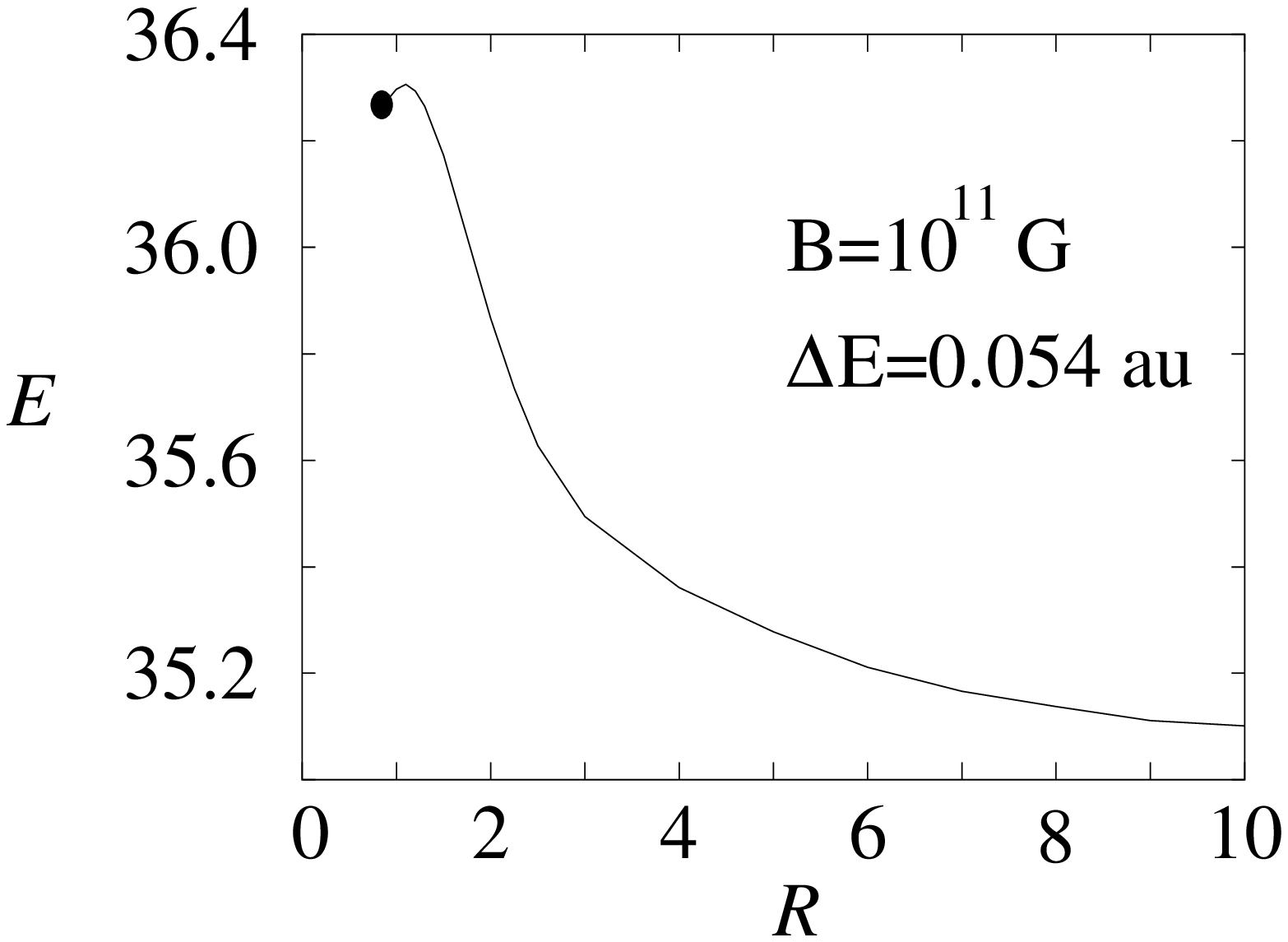,height=2.25in,angle=-90}}\\
{\psfig{figure=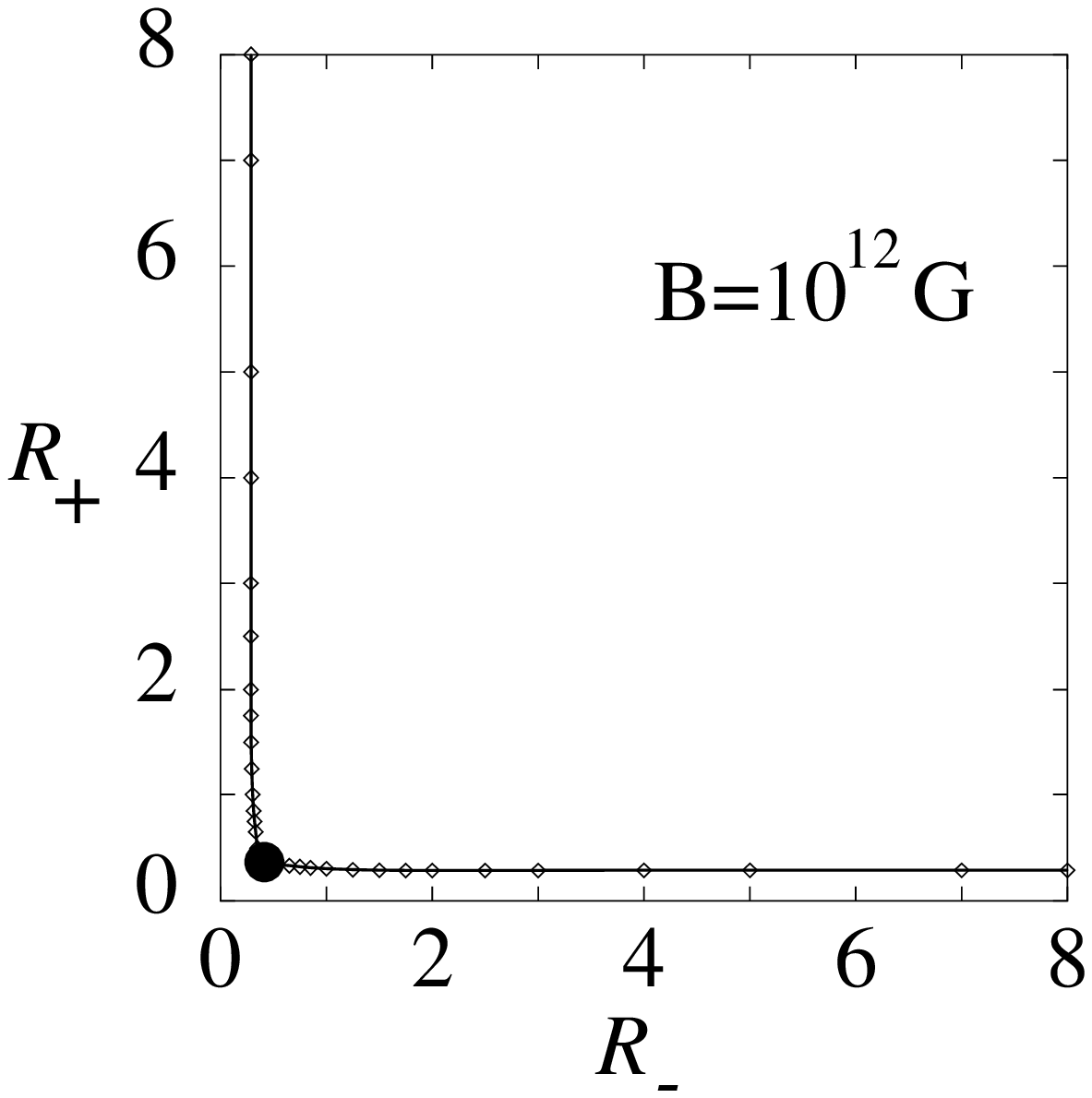,height=2.25in,angle=-90}} & &
{\psfig{figure=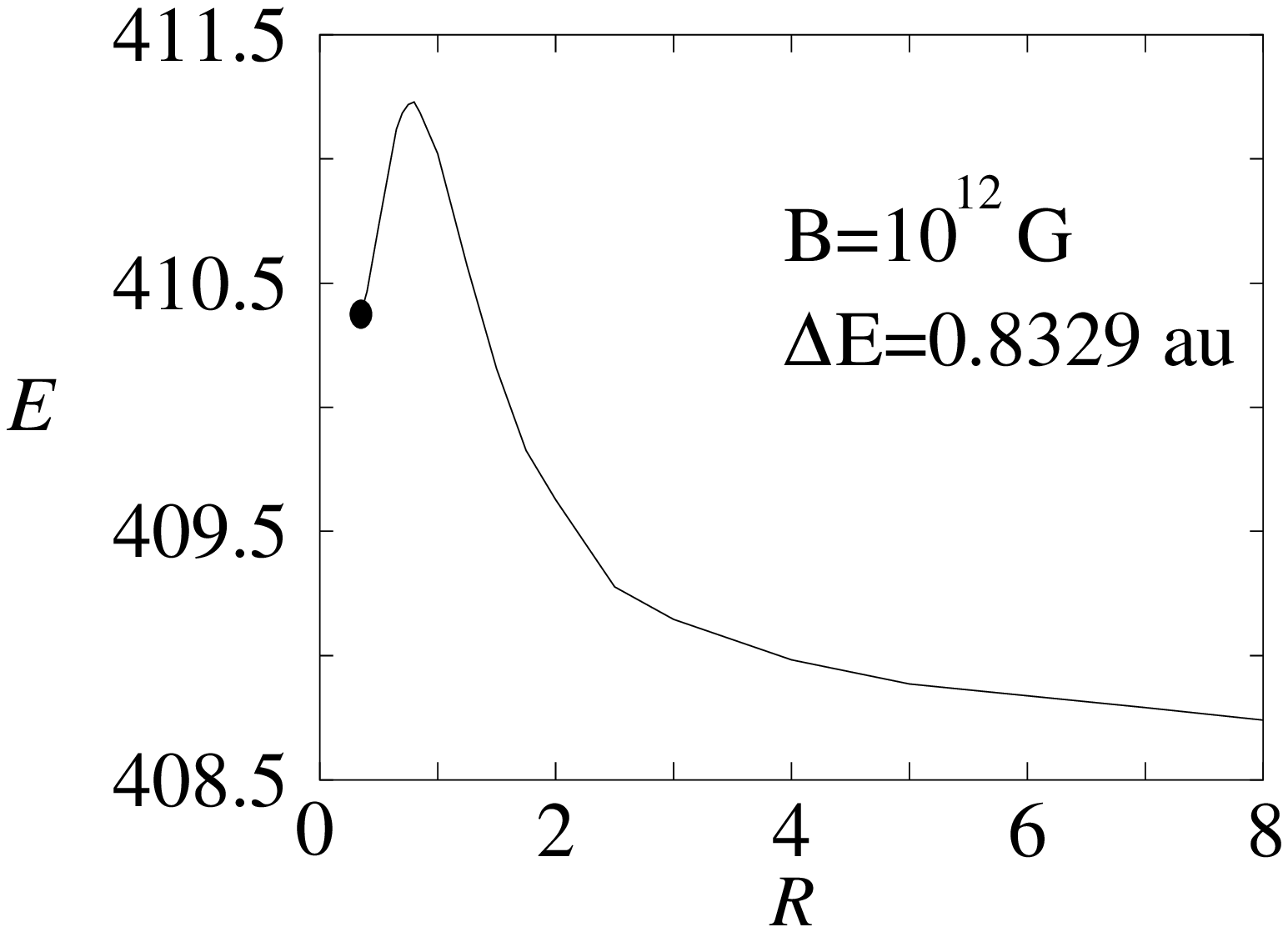,height=2.25in,angle=-90}}\\
{\psfig{figure=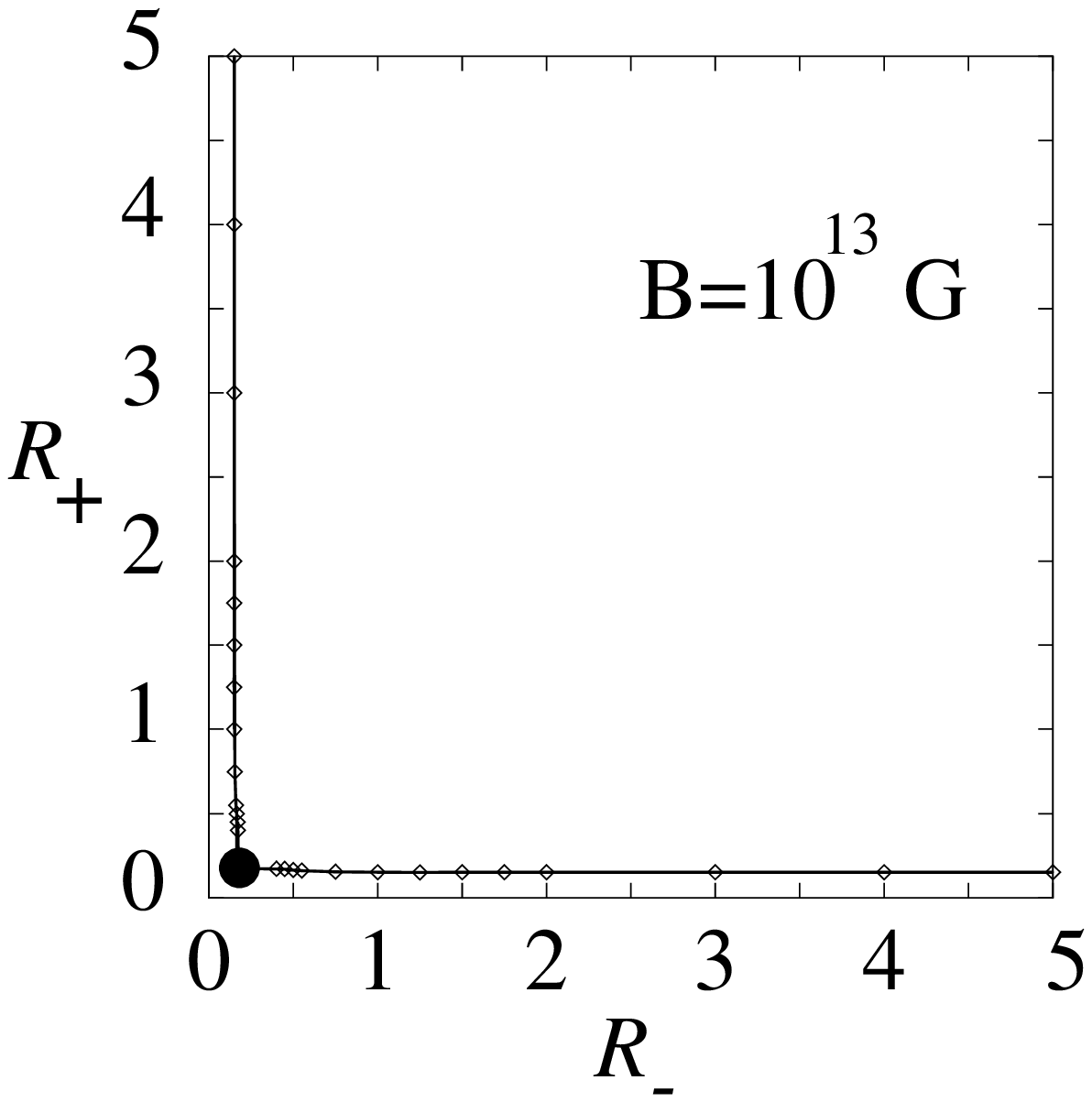,height=2.25in,angle=-90}} & &
{\psfig{figure=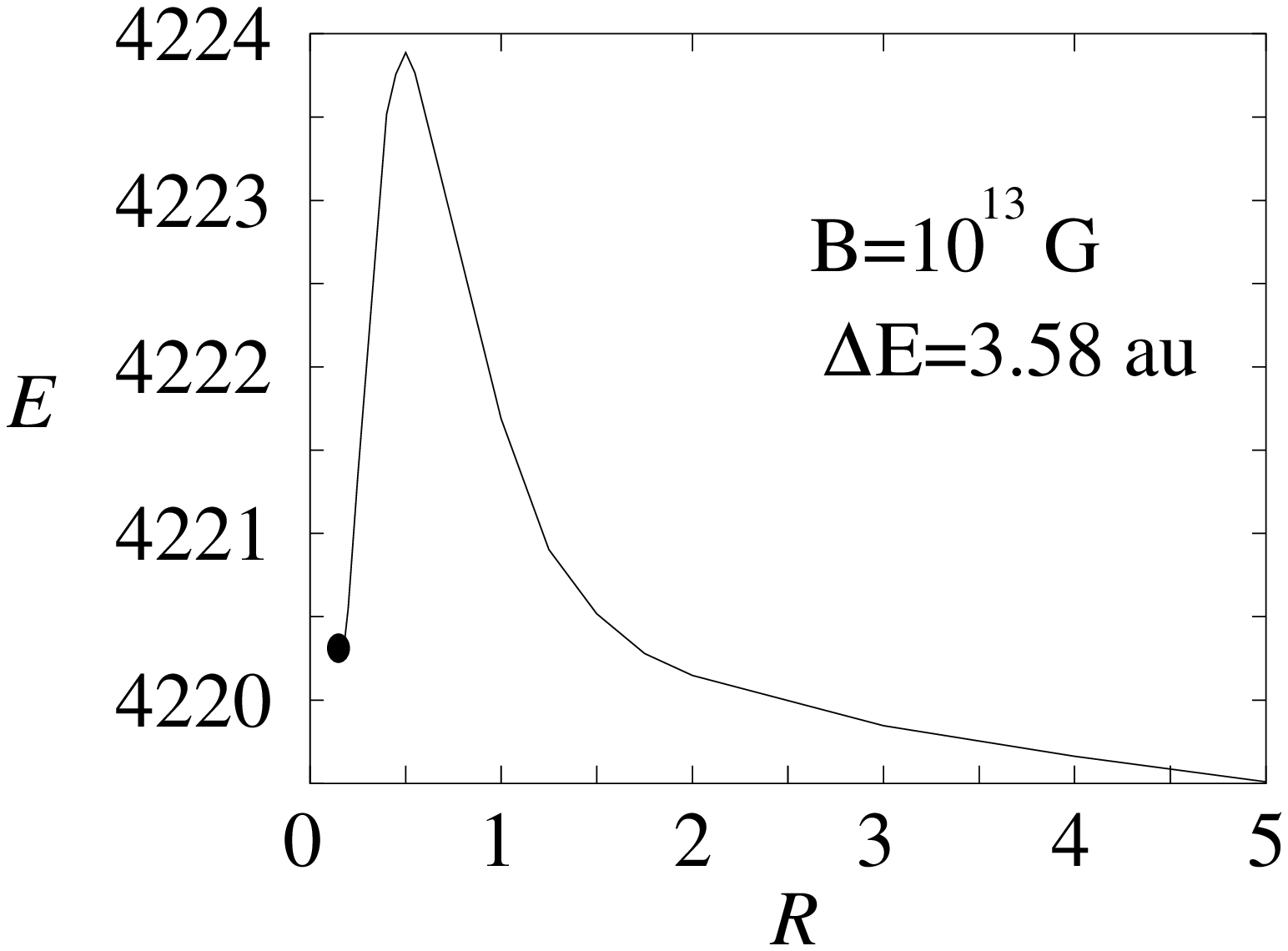,height=2.25in,angle=-90}}\\[5pt]
\hspace{15pt} {\large (a)} & & \hspace{28pt} {\large (b)}
\end{array}
\]
\caption{Valleys in the electronic potential energy surfaces,
  $E_{total}(R_+,R_-)$, (a) and their profiles (b) for various
  magnetic fields. The position of the minimum is indicated by a
  bullet, and $\Delta E$ denotes the depth of the well: the distance
  between top of the barrier and the value of minimum.}
\end{figure}

\end{document}